# Homogeneous nucleation mechanism of NaCl in aqueous solutions

Qiang Sun[*]

Key Laboratory of Orogenic Belts and Crustal Evolution, Ministry of Education, The School of

Earth and Space Sciences, Peking University, Beijing, 100871, China

**Abstract**

In this study, molecular dynamic simulations are employed to investigate the nucleation of NaCl crystal in solutions. According to the simulations, the dissolved behaviors of NaCl in water are dependent on ion concentrations. With increasing NaCl concentrations, the dissolved $Na^+$ and $Cl^-$ ions tend to be aggregated in solutions. In combination with our recent studies, the aggregate of dissolved solutes is mainly ascribed to the hydrophobic interactions. Thermodynamically, no barrier is needed to overcome in the formation of the solute aggregate. This is different from the two-step nucleation mechanism. In comparison with the classical nucleation theory (CNT), due to the aggregate of dissolved solutes, this lowers the barrier height of nucleation, and affects the nucleation mechanism of NaCl crystal in water.

---

[*] Corresponding author

E-mail: QiangSun@pku.edu.cn

## 1. Introduction

The nucleation of crystals in liquids is one of the most ubiquitous phenomena in nature. It is crucial in geology and environmental sciences, where the formation and behavior of crystalline solids may change with temperature, pressure, and chemical environment. In addition, nucleation plays an important role in many practical applications, ranging from semiconductors, metals, chemicals, to pharmaceutics. Therefore, many experimental and theoretical works have been devoted to investigate the mechanism of nucleation from aqueous solutions.

Nucleation is generally described by the classical nucleation theory (CNT) [1-5]. CNT is based on the assumption that the free energy necessary to create a nucleus of n particles can be divided into a favorable term, proportional to the number of particles in the nucleus, and an unfavorable term, proportional to the dividing surface between the nucleus and the solution. The free energy difference can be analytically expressed as,

$$\Delta G_{CNT} = -\Delta \mu \cdot n + \gamma \cdot s \qquad (1)$$

where $\Delta\mu$ is the difference in chemical potential between the crystal and the liquid phase, n is the number of molecules in the crystal phase, $\gamma$ is the surface tension, and s is the surface of the nucleus. Although CNT successfully describes various phenomena, it fails to predict homogeneous nucleation rates that are at least ten orders of magnitude slower than experimentally measured rates.

Recently, many experimental approaches [6-18] have been employed to investigate the thermodynamics and kinetics of crystal nucleation in liquids, such as optical microscopy, atomic force microscopy, cryo-TEM, ultrafast X-ray scattering, et al. These lead to the confirmation of the presence of the intermediate phase, such as amorphous calcium carbonate (ACC) [8,9], which are relevant to the solidification pathway of the dissolved solutes. Therefore, a two-step mechanism of nucleation of crystals in solution was put forth [19-21]. In the first step, amorphous nuclei are formed-their surface energy is lower than that of crystalline nuclei as a consequence of their disordered interfaces with the solution. In the second step, an amorphous-to-crystalline transition takes place in the middle of the amorphous phase. Compared with direct crystallization from solution, this transition has to overcome a much lower free energy barrier.

In addition, many works [22-28] have been conducted to investigate the nucleation mechanism of NaCl from the solutions. Zahn [22] found that ion aggregates were particularly stable when a $Na^+$ ion was octahedrally coordinated by $Cl^-$ ions. In Giberti et al. study [23], they employed metadynamics simulations to find an intriguing wurtzite-like polymorph, which was suggested to be an intermediate route from brine to the final rock salt structure. In addition, according to Chakraborty and Patey study [28], this means that a dense but unstructured NaCl nucleus is firstly formed, and rearranged into the rock salt structure. Therefore, this is in agreement with a two-step mechanism.

In this study, molecular dynamic simulations are employed to investigate the nucleation mechanism of NaCl crystal in solutions. According to the simulations, the dissolved behaviors of NaCl in water are dependent on ion concentrations. With increasing NaCl concentrations, the dissolved $Na^+$ and $Cl^-$ ions tend to be aggregated in solutions. In combination with our recent studies [29-35], this is mainly ascribed to the hydrophobic interactions. Due to the aggregate of dissolved solutes, this lowers the barrier height of nucleation, and affects the nucleation mechanism of NaCl crystal in water.

## 2. Molecular dynamics simulations

### 2.1 Simulated systems

Molecular dynamics (MD) simulations are carried out to investigate the nucleation of NaCl crystal in over-saturated solutions using the GROMACS (version 5.14) [36,37]. The simulated results are analyzed through Plumed (version 2.4) [38,39]. In principle, the nucleation mechanism may be closely related to the dissolved behaviors of NaCl in water. To investigate the structure of NaCl solutions, MD simulations are conducted on various solutions, such as $NaCl-H_2O$, $Cl^--H_2O$, and $Na^+-H_2O$ (Table 1).

In this study, The SPC/E water model was employed to represent the water molecules, and the OPLS force field was used to describe for the ion interaction parameters. The simulations were carried out in the NPT ensemble. The simulated temperature was kept at 300 K, employing Nose-Hoover thermostat dynamics. The pressure was maintained at 1 atm using a Parrinello-Rahman pressure coupling. The simulated box was initially kept at 42 Å×42 Å×42 Å, and NaCl salts were uniformly embedded in the box. Additionally, periodic boundary conditions were applied in all three directions. The Lennard-Jones interactions were truncated at 1.0 nm. The particle mesh Ewald method was used to calculate the long range electrostatics forces. Additionally, each simulation time was 60000 ps with a time step of 2 fs.

### 2.2 Order parameters

Generally, the Steinhardt parameters [40] are applied to measure the degree of order in the system. In this work, both $6^{th}$ order Steinhardt parameters (q6) and local Q6 parameters (LOCAL Q6) are calculated to measure the degree of order during nucleation of NaCl in the solutions.

Regarding to the 6$^{th}$ order Steinhardt parameters, it measures the degree to which the first coordination shell around an atom is ordered. With reference to atom i, the Steinhardt parameter are calculated as,

$$q_{6m}(i) = \frac{\sum_j \sigma(r_{ij}) Y_{6m}(r_{ij})}{\sum_j \sigma(r_{ij})} \qquad (2)$$

where $Y_{6m}$ is one of the 6$^{th}$ order spherical harmonics so m is a number that runs from -6 to +6. The function $\sigma(r_{ij})$ is a switching function that acts on the distance between atoms i and j. Additionally, for LOCAL Q6 parameter, it measures the extent to which the orientation of the atoms in the first coordination sphere of an atom matches the orientation of the central atom. It can be determined as,

$$s_i = \frac{\sum_j \sigma(r_{ij}) \sum_{m=-6}^{6} q_{6m}^*(i) q_{6m}(j)}{\sum_j \sigma(r_{ij})} \qquad (3)$$

where $q_{6m}(i)$ and $q_{6m}(j)$ are the 6$^{th}$ order Steinhardt vectors calculated for atom i and atom j, and the asterisk denotes complex conjugation.

### 2.3 Metadynamics

To investigate the nucleation mechanism from the solutions, the changes of free energy is reconstructed through Metadynamics method of Laio and Parrinello [41,42]. Metadynamics belongs to the family of enhanced sampling techniques in which the probability of visiting high free energy states is increased by adding to the Hamiltonian an adaptive external potential. The external repulsive potential is typically written as a series of Gaussian functions that are deposited in the space of collective variables (CVs) as,

$$V(S,t) = \sum_{kT<t} W(kT) \exp\left(-\sum_{i=1}^{d} \frac{(s_i - s_i^{(0)}(kT))^2}{2\sigma_i^2}\right) \quad (4)$$

where W and σ are the height and the width of the added Gaussians, respectively. Due to the external potentials, the system will freely diffuse above potential energy surface, completely sampling the CVs space. At this point, the free energy surface (FES) as a function of the set of CVs can be obtained as the negative of the repulsive bias deposited during the course of the simulation,

$$V(s) = -F(s) \quad (5)$$

### 3. Discussions

#### 3.1 The structure of NaCl solutions

The NaCl RDFs have a first contact maximum at 2.8 Å (CIP, contact ion pairs), second solvent-separated maximum at 5.1 Å (SSIP, solvent separated ion pairs), and a weak third maximum at about 7 Å corresponding to an attraction of fully hydrated ions. For dilute NaCl solutions (0.03 M), the RDFs are dominated by SSIP states (Figure 1). With increasing salt concentrations, this leads to a rise of the first maximum and a fall of the second in the Na-Cl RDFs (Figure 1). This is in correspondence with other studies on the structure of NaCl solutions [43,44]. Additionally, these changes can also be observed in the experimental measurements of aqueous LiCl solutions [45-47].

In principle, Radial Distribution Function $g_{ij}(r)$, where $r=|r_i-r_j|$ stands for the separation between a particle of component i and of component j, gives the probability of finding two particles at some distance r, taking account of density and geometric effects. This is related to the potential of mean force (PMF) between two particles, and can be expressed as,

$$W_{PMF(r)} = -k_B T \ln g(r) \qquad (6)$$

Therefore, the changes of $g_{ij}(r)$ indicate that the PMF of the separation between the particles is dependent on the NaCl concentrations.

For dilute NaCl solutions, the Na-Cl RDFs are dominated by SSIP. This means that the ions prefer to SSIP in dilute NaCl concentrations, and SSIP is thermodynamically stable than CIP. It is contrast with the PMF between a $Na^+$ and a $Cl^-$ ions in water [48], which shows that the first minimum (CIP at 2.8 Å) is lower than the second minimum (SSIP at 5.1 Å). Therefore, this indicates that the CIP and SSIP configurations may be separated by a high (several kT) effective potential barrier and transitions between them are rare events. In other words, it seems that there exists repulsive force between the dissolved ions in dilute NaCl solutions. Therefore, it is necessary to study the structure of water, and the effects of dissolved NaCl on water structure.

Many works have been devoted to investigate the structure of liquid water, which can roughly be partitioned into two categories: (a) mixture models and (b) continuum models [49]. Based on Raman spectroscopic studies on water structure [29-31], when three-dimensional hydrogen-bonded networks appear, various OH vibrational frequencies correspond to different hydrogen-bonded networks in the first shell of a water molecule (local hydrogen bonding). This indicates that OH vibrations are mainly dependent on the local hydrogen bonding of a water molecule, and the effects of hydrogen bonding beyond the first shell on OH vibrations are weak. Therefore, for ambient water, a water molecule interacts with neighboring water molecules (in the first shell) through various local hydrogen-bonded networks.

When NaCl salts are dissolved in water, they dissociate into ions that are hydrated. Due to the interactions between dissolved ions and water molecules around the ions, the dissolved ions affect the structure of hydrated water molecules, which may be different from that in bulk water. The OH vibrations are dependent on local hydrogen bondings, therefore the dissolved $Na^+$ and $Cl^-$ ions mainly affect the structure of water molecules within their first coordination shell. In fact, this can be demonstrated by the femtosecond mid-infrared spectroscopic studies [50,51], which means that the orientational dynamics of water is not modified outside the first solvation shell of dissolved ions.

In principle, when solutes are dissolved into water, the thermodynamic functions may contain solute-solute, solute-solvent and solvent-solvent interaction energies, respectively.

$$\Delta G = \Delta G_{Water-water} + \Delta G_{Solute-water} + \Delta G_{Solute-solute} \qquad (7)$$

In our Raman spectroscopic studies on NaCl solutions [30], as NaCl is dissolved into water, this mainly lowers the sub-band around 3220 $cm^{-1}$, and raises the sub-band around 3430 $cm^{-1}$. Therefore, the dissolved NaCl breaks the hydrogen bondings of water, and the strength of NaCl-water is weaker than that of water-water. Therefore, in dilute NaCl concentrations, the repulsive force between the $Na^+$ and $Cl^-$ ions may be closely related to hydrogen bonding of water, which results in the ions to tend to occupy SSIP configurations.

From the simulations, with increasing NaCl concentration, this increases the probability of CIP formation but decreases that of SSIP. This means that dissolved ions tend to form CIP rather than SSIP. Additionally, as increasing NaCl salts, these decrease the separations between $Cl^-$ and $Cl^-$ ions, and between $Na^+$ and $Na^+$ ions, especially $g_{Cl-Cl(r)}$ (Figure 2). This is in agreement with experimental studies on the structure of NaCl and LiCl solutions [43-47]. Therefore, it can be derived that the dissolved $Na^+$ and $Cl^-$ ions are not homogeneously distributed in water, and tend to be aggregated with increasing NaCl concentrations. This is different from the dissolved behaviors of $Na^+$ and $Cl^-$ ions in dilute concentrations.

In addition, the ion aggregation can also be demonstrated by the dependence of hydrogen bondings on NaCl concentrations (Figure 3). In this study, the geometrical definition of hydrogen bonding is utilized to determine the hydrogen bonds in water [52]. If $r_{OO}$ and $\angle OOH$ are less than 3.5 Å and 30°, a hydrogen bonding is considered to exist between two water molecules. With increasing NaCl concentrations, this decreases the hydrogen bonding number of water. This is due to the aggregate of dissolved ions in high salt concentrations, which decreases the surface area of ions available for water molecules.

In this study, the ion aggregated distribution in solutions is termed as the aggregate. For NaCl solutions, according to the calculated $g_{Na-Cl}(r)$, the ion is engaged in the aggregate as the separation between the $Na^+$ and $Cl^-$ being less than the first minimum (3.5 Å). Similar concepts are also used to reflect the aggregation, such as amorphous calcium carbonate (ACC) or pre-nucleation clusters (PNC) [8-11]. From this work, the ion aggregate can be found even in dilute NaCl solutions. With increasing ion concentrations, this increases the number of ion aggregation (Figure 4).

In fact, the solute aggregates have been reported in many experimental methods [6-18], such as transmission electron microscopy (TEM), atomic force microscopy (AFM), small angle X-ray scattering and light scattering. They are described to be short-range order and long-range disorder, and formed via aggregation of solution species comprised of ion complexes or multi-ion clusters, and may exist in equilibrium with the free monomers. The aggregate is observed not only for organic and colloidal systems but also for various electrolyte solutions. Additionally, besides the highly concentrated solutions, there is now much evidence indicating that ion aggregation also taking place at significantly lower concentrations (<1 M) [53,54]. Therefore, it is necessary to investigate the driving force of solute aggregation in water.

In the recent work by Marcus [55], the average distance apart of the centers of the ions in a $c$ M solution of a symmetrical electrolyte is d(nm)=0.94[c/M]. Therefore, if the dissolved ions are homogeneously distributed in water, ion pairing can be expected at highly concentrated (>1 M) solutions. Of course, this cannot be applied to understand the ion aggregations at significantly lower concentrations (<1 M). Regarding to the origin of driving forces for association between oppositely charged biological or synthetic polymers in aqueous solution, it is long identified as electrostatic force [54]. This attraction is broken down into an entropic component, due to loss of counterions, and an enthalpic component, stemming from Coulombic attraction between opposite charges.

To investigate whether the driving force of ion aggregation comes from electrostatic attractive force or not, the simulations are also conducted on single anion ($Cl^-$) and cation ($Na^+$) solutions where only repulsive forces exist between the ions. In comparison with RDFs of NaCl solutions, the Cl-Cl RDF of $Cl^-$-$H_2O$ and Na-Na RDF of $Na^+$-$H_2O$ systems are similar to the corresponding RDFs of NaCl solutions with the same solute numbers (Figure 2 inlet). In addition, with increasing $Cl^-$ and $Na^+$ concentrations, these also decrease the separations of $Cl^-$-$Cl^-$ and $Na^+$-$Na^+$ (Figure 2 inlet). Therefore, with increasing ion concentrations, the aggregate can also be expected in $Cl^-$-$H_2O$ and $Na^+$-$H_2O$ systems. Additionally, in comparison with Cl-Cl RDF of $Cl^-$-$H_2O$ and Na-Na RDF of $Na^+$-$H_2O$ systems, the corresponding RDFs of NaCl solutions slightly move to lower separations, which are undoubtedly related to electrostatic attractive force between $Na^+$ and $Cl^-$ ions. From these, it can be derived that the driving force of ion aggregation is mainly related to hydrogen bondings of water, which affects the global distributions of dissolved ions. Additionally, the local distributions may be affected by electrostatic force between $Na^+$ and $Cl^-$ ions.

To evaluate the role of hydrogen bonding of water, the hydrogen bonding number is also determined in the process of ion aggregations. In this study, the NaCl solutions are initially setup where the $Na^+$ and $Cl^-$ ions are homogeneously distributed in water. As the simulated system approaching the equilibrated state where the ions are aggregated, this is accompanied with the increase of hydrogen bonding in water (Figure 3 inlet). Because the strength of water-water hydrogen bonding is stronger than that of NaCl-water interactions, the driving force of ion aggregation can reasonably be ascribed to maximize the hydrogen bonding in water.

In our recent study [33], according to the structural studies on water and air/water interface [29-32], the hydration free energy is derived, which is utilized to investigate the physical origin of hydrophobic effects. It can be found that hydrophobic effects are ascribed to the structural competition between hydrogen bondings in bulk water and those in interfacial water [33]. With increasing solute concentrations, it can be divided into initial and hydrophobic solvation processes, which also corresponds to various dissolved behaviors of solutes, such as dispersed and accumulated. Of course, this is in agreement with the simulations on NaCl solutions.

In general, hydrophobic effects are described as the tendency of non-polar molecules or molecular surfaces to aggregate in an aqueous solution. From our recent study on the physical origin on hydrophobic effects [33], it is attributed to the strength of hydrogen bonding of bulk water is stronger than that of interfacial water. In other words, hydrophobic effects may be extended to other systems only if the strength of solute/solvent interface is weaker than that of bulk solvent. Therefore, hydrophobic interactions can be applied to understand the dissolved behaviors of NaCl salts in water.

As Na$^+$ and Cl$^-$ ions are dissolved into water, they mainly affect the hydrogen bonding of water molecules in the first shell. Because the strength of interfacial water is weaker than that of bulk water, ion aggregation can be expected as increasing NaCl concentrations. Of course, it should be noted that the tendency of ion aggregation may be counterbalanced by thermal motions. Additionally, in combination with our recent study on hydrophobic effects [33], it can be derived that the hydrophobic interactions of ion aggregation may be inversely proportional to the size of ion aggregation. Therefore, as increasing the size of ion aggregation, this also increases the strength of hydrophobic interactions.

**3.2 NaCl nucleation**

In this study, MD simulations are conducted on slightly oversaturated NaCl solutions (5.2322 M) to investigate the nucleation of NaCl crystal in solutions. Based on the calculated Na-Cl RDFs, these can be applied to study the NaCl nucleation from the solutions. In comparison with the Na-Cl RDFs before the nucleation, the second peaks are well split as NaCl crystal appears in the solutions (Figure 5). As NaCl continues to crystallize from the solutions, both the third and fourth peaks appear in the RDFs. Additionally, the nucleation can be found to first take place in the solute aggregate of the solutions. This is in accordance with other experimental and theoretical studies on the nucleation mechanism [20,21].

In some works, this nucleation is explained through Ostwald's Law of Stages or Ostwald's Rule [56]. It means that, when several solid phases exist, the formation of the thermodynamically stable phase can be preceded by metastable intermediates that stepwise transform to the final product. Namely that, in the case of a compound capable of crystallizing in several forms, it will be the least stable form which is first produced by spontaneous crystallization, followed successively by forms of increasing stability. However, the aggregate is found in not only concentrated but also dilute solutions, the structure can be described as short-range order and long-range disorder. Therefore, it is unreasonable to understand the origin of the aggregate through Ostwald's rule.

Different from the aggregate, the structure of crystal is characterized by the periodic lattice of elementary unit. In this work, LQ6 is calculated and utilized to distinguish the NaCl nucleation from the aqueous solutions. As the NaCl crystal nucleates from the solutions, this increases the order parameter of LQ6 (Figure 6). This is also in correspondence with the changes of Na-Cl RDFs. In addition, the nucleation of NaCl crystal can also be observed in the changes of coordination number of the first shell. Therefore, both LQ6 and the first coordination number can be applied to measure the nucleation of NaCl crystal in the solutions.

To investigate the changes of free energy in the process of NaCl nucleation, based on LQ6 and first coordination number, the free energy surface (FES) can be determined through METAD method (Figure 7). It can be found that only one potential barrier is necessary to overcome so that the nucleation of NaCl crystal proceeds. Based on the calculated LQ6 of the solutions, the lowest zone of FES is in correspondence with the formation of ion aggregation in the solutions. Therefore, no barrier is necessary to overcome in the formation of ion aggregate. Due to the ion aggregation, this will make the system to be more thermodynamically stable. In other words, the dissolved ions tend to be aggregated in the solutions.

From the simulations, the nucleation of NaCl crystal first takes place in the ion aggregate. This indicates that the ion aggregation should be helpful to the NaCl nucleation from the solutions. In other words, the ion aggregate may correspond to the nucleation site with relatively lower height of nucleation barrier. In combination with the above discussion, it can be derived that, due to hydrophobic interactions, this leads to the formation of solute aggregate, and also lowers the height of nucleation barrier.

According to the discussion on the structure of NaCl solutions, it can be found that the ion aggregate is not considered in CNT theory. To understand the nucleation mechanism from aqueous solutions, it is necessary to take into account the hydrophobic interactions. Because the ion aggregate lower the height of potential barrier, the CNT can reasonably be revised as,

$$\Delta G_{Rev} = \Delta G_{CNT} - \Delta G_H \quad (8)$$

where $\Delta G_H$ means the hydrophobic interactions in the formation of ion aggregate. Different from the critical nuclei of CNT, the critical aggregate can be expected, $Agg_C$, which corresponds to the aggregate as the nucleation of solute occurs in solutions.

In combination with our recent studies on hydrophobic interaction [33-35], the $\Delta G_H$ is the difference of Gibbs energy as the solutes are transformed from dispersed to accumulated distributions in water. In this work, the solute is regarded as an ideal hydrophobic sphere. Regarding to a sphere aggregate, if the solute number of aggregate is n, the $\Delta G_H$ is expressed as,

$$\Delta G_H = 8 \cdot \frac{\left(n - n^{-1/3}\right)}{r} \cdot \Delta G_{DDAA} \cdot r_{H_2O} \quad \left(R = n^{1/3} \cdot r\right) \quad (9)$$

where $\Delta G_{DDAA}$ is the Gibbs energy of DDAA hydrogen bonding, $r_{H2O}$ is the radius of a $H_2O$ molecule, n is the atomic number of aggregate, r is the radius of solute, and R is the size of solute aggregate. From this, it can be derived that the larger of the atomic number of aggregate (or aggregate size), the stronger of hydrophobic interactions. Therefore, the nucleation is expected to take place in the largest aggregate, and the $Agg_C$ corresponds to the largest aggregate in the nucleation process (Figure 8). In fact, this may be applied to understand why nucleation occurs in supersaturated solutions. In addition, this can also be extended to investigate the mechanism of heterogeneous nucleation from the solutions. This may be covered in our next work.

From this work, the nucleation of NaCl crystal can be described as, the ion dispersed distribution → the ion aggregate → the nucleation. It should be noted that, different from two-step nucleation, only one barrier is needed to overcome in the process of solute nucleation. In two-step mechanism, it is also a potential barrier to overcome to form the solute aggregate. In fact, regarding to the formation of solute aggregate, it is closely related to hydrophobic interactions, and no barrier is necessary to overcome. In other words, the driving force of ion aggregate is due to maximize the hydrogen bonding of bulk water. In addition, due to the solute aggregate, this lowers the height of potential barrier of nucleation.

Regarding for the CNT theory, it does not take into account the effects of hydrophobic interactions on the dissolved behaviors of solutes in water. Therefore, CNT may be applied to investigate the nucleation as hydrophobic interactions can be ignored. In combination with this study, this means that CNT can be utilized to understand the nucleation mechanism as the solutes are dispersed in solutions.

**4. Conclusions**

In this work, molecular dynamic simulations are applied to investigate the nucleation mechanism of NaCl in solutions. From this study, the following conclusions can be derived,

(1) According to the simulations, the dissolved behaviors of NaCl in water are dependent on ion concentrations. With increasing NaCl concentrations, the dissolved behaviors of $Na^+$ and $Cl^-$ ions are transformed from dispersed to aggregated distributions in water.

(2) In combination with our recent studies on hydrophobic interactions, the solute aggregate is mainly ascribed to the hydrophobic interactions. Thermodynamically, no barrier is needed to overcome in the formation of the solute aggregate. Therefore, this is different from the two-step nucleation mechanism.

(3) In comparison with the classical nucleation theory (CNT), due to the formation of ion aggregate in solutions, this lowers the barrier height of nucleation. Therefore, the nucleation of crystal can be expected to take place in the largest aggregate, and the $Agg_C$ is expected during the nucleation.

## Acknowledgements

This work is supported by the National Natural Science Foundation of China (Grant Nos. 41773050).

**Table 1.** The simulated systems in this work.

**Figure 1.** The calculated $g_{Na-Cl(r)}$ of various NaCl solutions. The inlet shows the dependence of the ratio of second peak to the first peak of $g_{Na-Cl(r)}$ on the NaCl concentrations.

**Figure 2.** (a) The calculated $g_{Cl-Cl(r)}$ functions of NaCl solutions. The inlet shows the $g_{Cl-Cl(r)}$ of $Cl^--H_2O$ systems and NaCl solutions. (b) The calculated $g_{Na-Na(r)}$ functions of NaCl solutions. The inlet shows the $g_{Na-Na(r)}$ of $Na^+-H_2O$ systems and NaCl solutions.

**Figure 3.** The dependence of hydrogen bondings of water on NaCl concentrations. The fitted line is shown in solid. The inlet shows the changes of hydrogen bonding of NaCl solutions ($NaCl:H_2O=10:2000$) as the system approaching the equilibrium.

**Figure 4.** The number of $g_{Na-Cl(r)}$ less than the first minimum (3.5Å) for various NaCl solutions. The fitted line is shown in solid.

**Figure 5.** The $g_{Na-Cl(r)}$ of NaCl solutions ($NaCl:H_2O=223:2000$) in the process of nucleation. The configurations are also shown.

**Figure 6.** The changes of local Q6 parameter of NaCl solutions (NaCl:$H_2$O=223:2000) during the nucleation. As the nucleation of NaCl crystal appears, the max aggregate of NaCl is also shown (atomic number is 169).

**Figure 7.** The FES as the function of LQ6 and $1^{st}$ coordination number in the process of nucleation from NaCl solutions (NaCl:$H_2$O=223:2000).

**Figure 8.** The homogeneous nucleation mechanism of dissolved solutes in water. Due to the formation of solute aggregate, this decreases the barrier of nucleation.

Table 1. The simulated systems in this work.

| No | System | Solute | $H_2O$ | Concentrations (M) |
|---|---|---|---|---|
| 1 | NaCl-$H_2O$ | 1 | 2000 | 0.02749 |
| 2 | NaCl-$H_2O$ | 2 | 2000 | 0.05505 |
| 3 | NaCl-$H_2O$ | 3 | 2000 | 0.08316 |
| 4 | NaCl-$H_2O$ | 4 | 2000 | 0.11143 |
| 5 | NaCl-$H_2O$ | 5 | 2000 | 0.13694 |
| 6 | NaCl-$H_2O$ | 10 | 2000 | 0.2762 |
| 7 | NaCl-$H_2O$ | 20 | 2000 | 0.54334 |
| 8 | NaCl-$H_2O$ | 30 | 2000 | 0.80641 |
| 9 | NaCl-$H_2O$ | 40 | 2000 | 1.08475 |
| 10 | NaCl-$H_2O$ | 50 | 2000 | 1.3422 |
| 11 | NaCl-$H_2O$ | 223 | 2000 | |
| 12 | $Cl^-$-$H_2O$ | 20 | 2000 | |
| 13 | $Cl^-$-$H_2O$ | 40 | 2000 | |
| 14 | $Na^+$-$H_2O$ | 20 | 2000 | |
| 15 | $Na^+$-$H_2O$ | 40 | 2000 | |

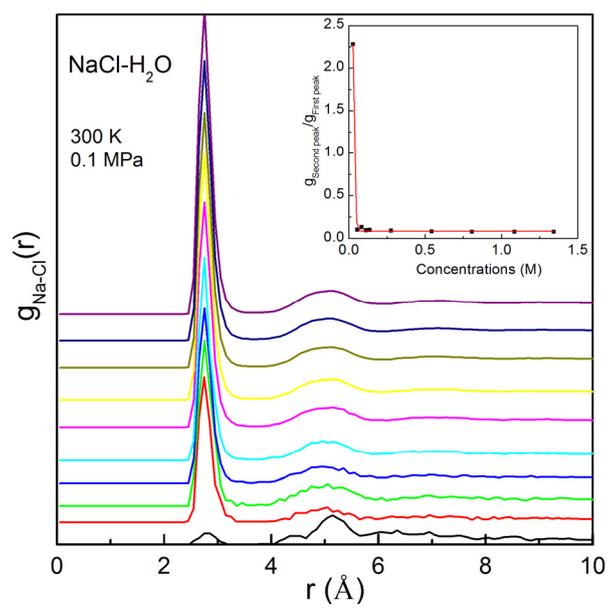

**Figure 1**

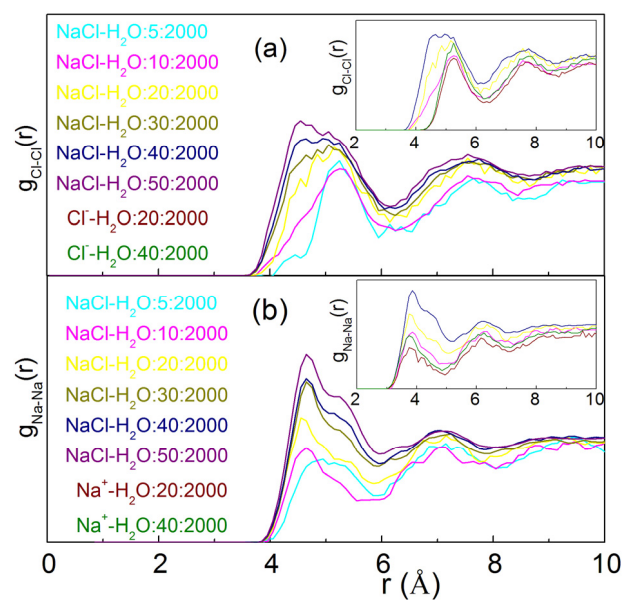

**Figure 2**

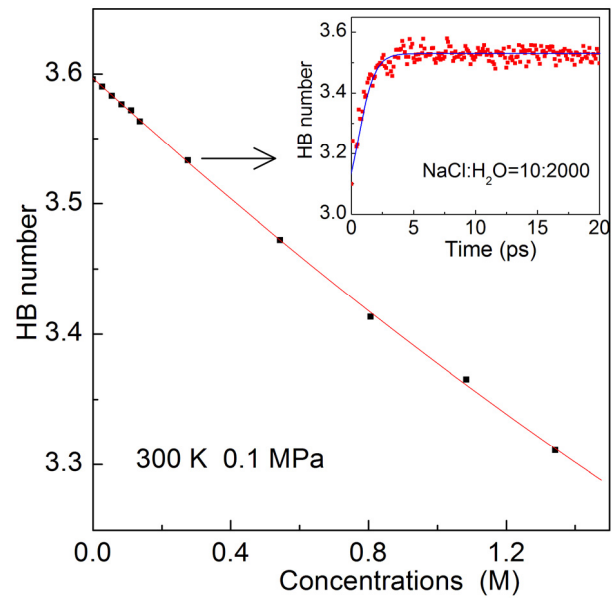

**Figure 3**

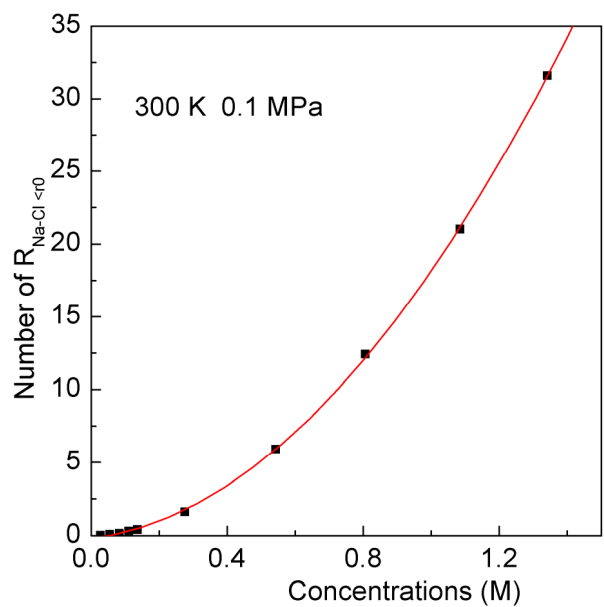

**Figure 4**

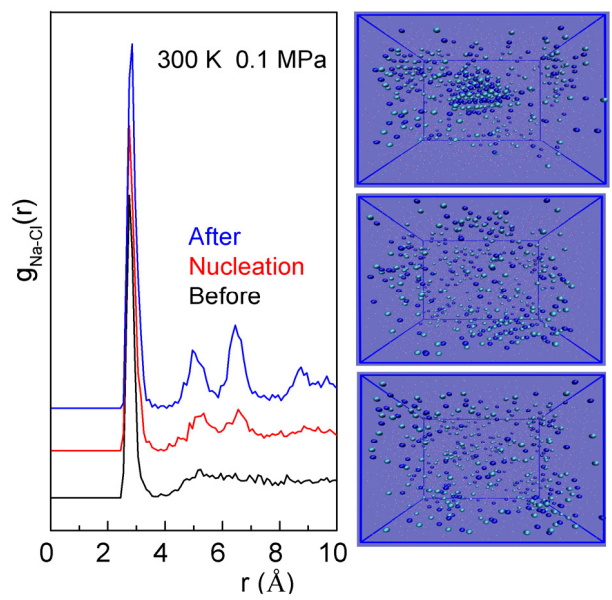

**Figure 5**

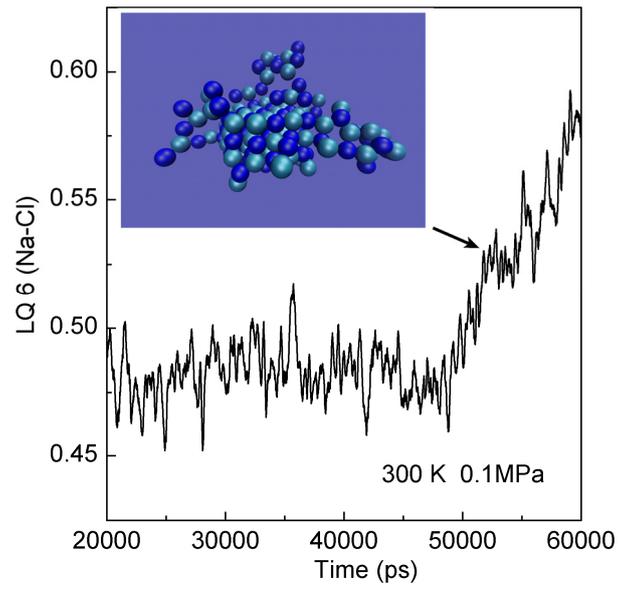

**Figure 6**

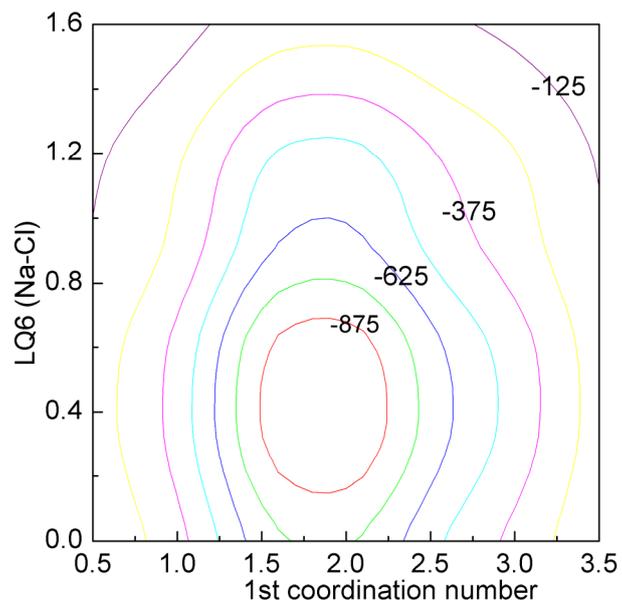

**Figure 7**

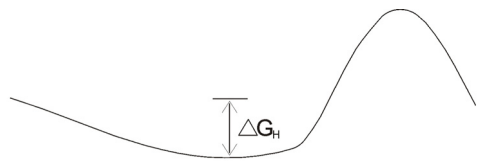

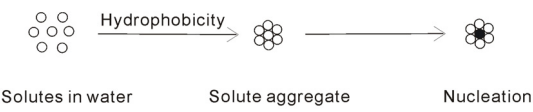

$$\Delta G_H = 8 \cdot \frac{\left(n - n^{-1/3}\right)}{r} \cdot \Delta G_{DDAA} \cdot r_{H_2O} \quad \left(R = n^{1/3} \cdot r\right) \begin{array}{l} r\text{ -the radius of hydrophobic solute} \\ n\text{-the solute number of aggregate} \\ R\text{-the radius of aggregate} \end{array}$$

$$\Delta G_{Rev} = \Delta G_{CNT} - \Delta G_H \longrightarrow \text{Agg}c \text{ (Critical aggregate, the largest aggregate as the nucleation occurs)}$$

Rev-CNT

**Figure 8**